\RequirePackage{fix-cm}
\documentclass[smallextended]{svjour3}       
\smartqed  
\usepackage{graphicx}
\usepackage{epstopdf}
\usepackage{amsmath,amssymb,mathrsfs}
 \journalname{my journal}
\begin{document}

\title{Symplectic Noise \& The Classical Analog of the Lindblad Generator
}
\subtitle{Does the Regression Hypothesis also fail in Classical Physics?}


\author{John E. Gough
}


\institute{J.E. Gough \at
              Aberystwyth University,
	Aberystwyth,
  SY23 3BZ,
  United Kingdom \\
              Tel.: +123-45-678910\\
              Fax: +123-45-678910\\
              \email{jug@aber.ac.uk}           
}

\date{Received: date / Accepted: date}

\maketitle

\begin{abstract}
We introduce the concept of Poisson Brackets for classical noise and, in particular, for 
pairs of canonically conjugate Wiener processes (symplectic noise). Phase
space diffusions driven by these processes are considered and the general
form of a stochastic process preserving the full (system and noise) Poisson
structure is obtained. Making the requirement that the classical stochastic model 
preserves the joint system and noise Poisson Brackets, we show that the model has actually
much in common with quantum markovian models. 	
\end{abstract}

\section{Introduction}
\label{intro}

In 1931 Onsager proposed a markovian dynamical model for the fluctuations of a classical dissipative system close to equilibrium in order 
to derive the celebrated reciprocal relations \cite{Onsager}. A more general theory was given by Casimir \cite{Casimir} who included the time-derivatives 
of the extensive variables of Onsager's theory as new Markov state variables. Subsequently, Onsager and Machlup added the further assumption of Gaussianity
\cite{Onsager_Machlup} to derive the regression theorem. We mention briefly that this work is in the time-domain, as opposed to the fluctuation-dissipation 
theory of Nyquist \cite{Nyquist} which is frequency domain. In the quantum version, there are known to be problems reconciling the Onsager and Nyquist theories,
often interpreted as a failure of the quantum regression theorem \cite{Talkner}.

It is well known that the generator of a quantum dynamical semigroup (QDS)
of superoperators takes the GKS-Lindblad form \cite{GKS,Lindblad}
\begin{eqnarray*}
\mathcal{L}X = \frac{1}{i\hbar} [ X , H] + \dfrac{1}{2} \sum_k \bigg\{ L^\dag_k [X,L_k ] + [L^\dag_k , X] L_k \bigg\} ,
\end{eqnarray*}
with $H$ self-adjoint. Lindblad \cite{Lindblad} characterized the generators by their dissipation $\mathcal{D}_{\mathcal{L}}$ which he defined to be 
\begin{eqnarray*}
\mathcal{D}_{\mathcal{L}} (X,Y) \triangleq
\mathcal{L} (X^\dag Y) - \mathcal{L}(X^\dag)\, Y -X^\dag \, \mathcal{L}Y,
\end{eqnarray*}
and showed that complete positivity implied dissipativity, that is $\mathcal{D}_{\mathcal{L}} (X,X)  = \sum_k [L_k , X ]^\dag [L_k , X ] \geq0$ (in fact, that it implied the stronger condition of complete dissipativity).
We may obtain the generator from a quantum stochastic model \cite{HP,Partha} with quantum stochastic differential equation (QSDE) defining a
unitary process
\begin{eqnarray*}
dU_t = \bigg\{ \sum_k L_k dB_k^\dag (t) -  \sum_k L_k^\dag dB_k (t) - ( \frac{1}{2}
\sum_k L_k^\dag L_k +iH)dt \bigg\} \, U_t
\end{eqnarray*}
where $B_k (t)$ are quantum annihilator processes living on a Fock space
describing the environment, and we have the nontrivial quantum It\={o}
relation $dB_j (t) \, dB_k (t)^\dag = \delta_{jk} \, dt$. In particular, for an
observable $X$ of the system, we set $j_t (X) = U^\dag_t (X \otimes I_{%
\mathrm{Fock}}) U_t$ and find
\begin{eqnarray*}
d j_t (X) = j_t (\mathcal{L} X) \, dt 
+ \sum_k j_t ([X,L_k]) \, dB_k^\dag (t) + \sum_k j_t ([L_k^\dag ,X ]) \, dB_k (t) .
\end{eqnarray*}
Taking the partial trace of $j_t (X)$ over the Fock vacuum leads to the QDS
generated by $\mathcal{L}$. We note that the noise can be written in terms of the two \textit{quadratures}
\begin{equation*}
Q_k (t) = B_k (t) +B_k(t)^\dag , \quad P_k (t) = \frac{1}{i} \left( B_k (t)
- B_k(t)^\dag \right) .
\end{equation*}
Separately, these are Wiener processes for the Fock vacuum state, but they
in fact are non-commuting operator-valued processes with $[Q_j (t) , P_k (s) ] = \delta_{jk} \mathrm{min}
(t,s) $, and
\begin{equation*}
(dQ_k)^2 = (dP_k)^2 =dt, \quad dQ_k \, dP_k = -dP_k \, dQ_k =i \, dt.
\end{equation*}

The quantum stochastic model essentially achieves a form of fluctuation-dissipation balance
at the dynamical level, which in the case of quantum theory is algebraic in nature. The problem of
including thermal noise, properties of detailed balance, whether there is a regression theorem, the relationship
with the Nyquist theory, etc., has been addressed extensively \cite{ALV,GZ}.
The quantum noise however becomes essentially classical if at most only one of the
processes in each quadrature noise pair $(Q_k,P_k)$ appears: more generally, we could add in classical jump processes along with the Wiener noise, 
and the class of generator for quantum systems with essentially classical noise is known \cite{KM} and is smaller
that the GKS-Lindblad class. It is therefore the case in quantum Markovian evolutions that
both the system and the noise are quantum!

It is at this point we return to the classical theory mechanics, and to the programme initiated by Onsager. Classical mechanics
has a geometric structure characterized by the Poisson Brackets. We argue that the general ``de-quantized'' version of quantum Markov stochastic models should capture the commutation relations of the noise as well as the system: that is, there is a Poisson Brackets for both the system and noise processes!

\section{Classical Noise}

We recall the classical phase space setting with
global canonical coordinates $(q,p)$ and standard Poisson Brackets
\begin{eqnarray*}
\{ f,g \} = \frac{\partial f\left( q,p\right)   }{\partial q}
\frac{\partial g\left(   q,p\right)   }{\partial p} 
-\frac{\partial g\left(   q,p\right)  }{\partial q}
\frac{\partial f\left(  q,p\right)   }{\partial p}.
\end{eqnarray*}
The restriction to one mechanical degree of freedom is for convenience only.

\subsection{Deterministic Flows on Phase Space}
Let us consider a deterministic dynamical flow on phase space given by
\begin{equation}
\dot{q}_{t}=v^{q}(q_{t},p_{t}),\;\dot{p}_{t}=v^{p}(q_{t},p_{t})
\end{equation}
and we have the local solutions $q_{t}=q_{t}\left( q,p\right) $ and $%
p_{t}=p_t \left( q,p\right) $ integral to the velocity vector field $v=\left(
v^{q},v^{p}\right) $ and with initial phase point $\left( q,p\right) $ at $%
t=0$. Then for any function $f$ on phase space we have
\begin{eqnarray}
\frac{d}{dt}f(q_{t},p_{t})=\left. v.\nabla (f)\right| _{(q_{t},p_{t})}.
\label{eq:diff_v}
\end{eqnarray}
From a geometric viewpoint, $v.\nabla $ is of course a tangent vector field on the phase space. In what follows, we could take the phase space to
be a general cotangent bundle with the canonical symplectic structure, or more generally a Poisson manifold \cite{Bloch}, however we restrict to
the standard case for transparency.

The Poisson Bracket of the evolved functions is
\begin{eqnarray*}
\left\{ f\left( q_{t},p_{t}\right) ,g\left(
q_{t},p_{t}\right) \right\} &=&\frac{\partial f\left(
q_{t}\left( q,p\right) ,p_{t}  \left( q,p\right) \right) }{\partial q}
\frac{\partial g\left( q_{t}\left( q,p\right) ,p_{t}\left( q,p\right) \right) }{\partial p}\\
&&-\frac{\partial g\left( q_{t}\left( q,p\right) ,p_{t}\left(
q,p\right) \right) }{\partial q}
\frac{\partial f\left( q_{t}\left(
q,p\right) ,p_{t}\left( q,p\right) \right) }{\partial p}.
\end{eqnarray*}
In what follows we wish to consider flow semi-groups generated by differential operators $\mathcal{L}$ other than vector fields. 
In particular, we have in mind the generators of Markov diffusions, which are second-order differential operators. 
\bigskip

\textbf{Definition:} We say that the flow is \textit{canonical} if we have 
\begin{eqnarray}
\left\{ f\left( q_{t},p_{t}\right) ,g\left( q_{t},p_{t}\right) \right\}
=\left. \left\{ f,g\right\} \right| _{\left( q_{t},p_{t}\right) },
\label{eq:Canonical}
\end{eqnarray}
for ever pair of smooth functions $f,g$.

\bigskip

For example, the linearly damped harmonic oscillator is
\begin{equation}
v^{q}_{\mathrm{DHO}}=\frac{p}{m},\quad v^{p}_{\mathrm{DHO}} =-m\omega ^{2}q-\gamma p  \label{eq:DHO}
\end{equation}
and we find $\nabla .v=-\gamma $, constant. For this case we see that the
Poisson Brackets are $\left\{ f\left( q_{t},p_{t}\right) ,g\left(
q_{t},p_{t}\right) \right\} =e^{-\gamma t}\,\left\{ f,g\right\} $.

As the classical analog of Lindblad's dissipation, we make the following definition.
\bigskip

\textbf{Definition: }Let $\mathcal{L}$ be a differential operator on phase
space. Its \textit{dissipation} is the object $\mathscr{D}_{\mathcal{L}}$
defined by
\begin{equation}
\mathscr{D}_{\mathcal{L}}(f,g)\triangleq \mathcal{L}\left\{ f,g\right\}
-\left\{ \mathcal{L}f,g\right\} -\left\{ f,\mathcal{L}g\right\}
\label{eq:dissip}
\end{equation}
for arbitrary twice-differentiable functions $f,g$. 

\bigskip
Some commentary is appropriate here. The classical definition of dissipation involves the Poisson Brackets, and so differs from the obvious analog of Lindblad's definition which would be $\Gamma_{\mathcal{L}}(f,g)\triangleq \mathcal{L}\left( f g\right)-  \mathcal{L} (f) g - f \mathcal{L}(g)$. The $\Gamma_{\mathcal{L}}$ operator is the well-known \emph{squared-field} operator from the geometric analysis of Markov generators \cite{BGL}. Classically we have $\Gamma_{\mathcal{L}} \equiv 0$ if and only if $\mathcal{L}$ is a tangent vector field (that is, a first order differential operator). In this sense $\Gamma_{\mathcal{L}}$ is a natural  analog of the Lindblad dissipation in the sense that they both measure how far $\mathcal{L}$ is away from being a derivation on the corresponding algebras, namely the algebra of smooth functions classically and the C*-algebra of operators on the Hilbert space in the quantum case. However, the product of operators encodes kinematic information, for instance the Heisenberg canonical commutation relations $qp-pq=i \hbar$, and our definition gives the appropriate physical definition of dissipation in the classical setting. The next two propositions single out Hamiltonian systems as those having no dissipation.

\textbf{Proposition 1} \textit{The dissipation associated with a vector field%
} $v$ \textit{on phase space is}
\begin{equation*}
\mathscr{D}_{v.\nabla }(f,g)=-(\nabla .v)\,\{f,g\}.
\end{equation*}

\bigskip

\textbf{Definition:} The dynamics is said to be \textit{Hamiltonian} if $%
v^{q}(q,p)=\frac{\partial H}{\partial p}$,  $ v^{p}(q,p)=-\frac{\partial H}{%
\partial q},$for some function $H$ on phase space. (Equivalently, $v.\nabla \equiv \{\cdot ,H\}$.) 

\bigskip

\textbf{Proposition 2} \textit{The dissipation associated with a vector
field } $v$ \textit{on phase space with differentiable components }$v^{q}$
\textit{\ and} $v^{p}$\textit{will vanish if and only if it is Hamiltonian.}

\bigskip

\textit{Proof:} By Proposition 1, we have $\mathscr{D}_{v}\equiv 0$ only if $%
\nabla .v=\frac{\partial }{\partial q}\left( v^{q}\right) +\frac{\partial }{%
\partial p}\left( v^{p}\right) =0$ at every point of phase space. It is
well-known from potential theory that this will then imply that $v^{q}=\frac{%
\partial H}{\partial p},v^{p}=-\frac{\partial H}{\partial q}$ for some
function $H$. Conversely, if $v$ is Hamiltonian then $\nabla .v=\frac{%
\partial }{\partial q}\left( \frac{\partial H}{\partial p}\right) +\frac{%
\partial }{\partial p}\left( -\frac{\partial H}{\partial q}\right) =0$,
which is of course Liouville's theorem. $\square $

Note that the converse also follows from the Jacobi identity for Poisson Brackets: $\mathscr{D}_{\left\{ \cdot
,H\right\} }(f,g) =\left\{ \left\{ f,g\right\} ,H\right\} -\left\{ \left\{
f,H\right\} ,g\right\} -\left\{ f,\left\{ g,H\right\} \right\}=0$.

\bigskip

\textbf{Proposition 3} \textit{The flow generated by a velocity field }$v$%
\textit{\ on phase space is canonical if and only if }$v$\textit{\ is
Hamiltonian.}

\bigskip

\textit{Proof:} Using (\ref{eq:diff_v}), we see that the infinitesimal form
for the canonical criterion is $v.\nabla \{f,g\}=\{v.\nabla
f,g\}+\{f,v.\nabla g\}$ for all $f$ and $g$, and this requires $\mathscr{D}%
_{v.\nabla }\equiv 0$. However, being dissipation-free is equivalent from
Proposition 2 to being Hamiltonian. $\square $

\subsection{Stochastic Flows on Phase space}

We now consider a stochastic diffusion on phase space driven by
independent Wiener processes $\{ Q_{k}\left( t\right) :$ $k=1,\cdots
,n,t\geq 0\} $ and satisfying the stochastic differential equations
(SDE)
\begin{eqnarray}
dq_{t} &=&v^{q}\left( q_{t},p_{t}\right) \,dt+\sum_{k}\sigma _{k}^{q}\left(
q_{t},p_{t}\right) \,dQ_{k}\left( t\right) ,  \notag \\
dp_{t} &=&v^{p}\left( q_{t},p_{t}\right) \,dt+\sum_{k}\sigma _{k}^{p}\left(
q_{t},p_{t}\right) \,dQ_{k}\left( t\right) .  \label{eq:SDE}
\end{eqnarray}
The SDEs are understood in the It\={o} sense, so all increments are future
pointing. The vector fields $\sigma _{k}=\left( \sigma
_{k}^{q}(q,p),\sigma _{k}^{p}(q,p)\right)$, for $k=1,\cdots ,n$,
fix the strength of the fluctuations. For a function $f$ on phase space, we find
\begin{equation*}
df\left( q_{t},p_{t}\right) =\left. \left( \mathcal{L}f\right) \right|
_{\left( q_{t},p_{t}\right) }\,dt+\sum_{k}\left. \left( \sigma _{k}.\nabla
f\right) \right| _{\left( q_{t},p_{t}\right) }\,dQ_{k}\left( t\right)
\end{equation*}
where the generator $\mathcal{L}$ of the diffusion is the second order
differential operator
\begin{equation*}
\mathcal{L}=v^{q}\,\frac{\partial }{\partial q}+v^{p}\,\frac{\partial }{%
\partial p}+\frac{1}{2}g^{qq}\,\frac{\partial ^{2}}{\partial q^{2}}+g^{qp}\,%
\frac{\partial ^{2}}{\partial q\partial p}+\frac{1}{2}g^{pp}\,\frac{\partial
^{2}}{\partial p^{2}}
\end{equation*}
and the diffusion coefficients are $g^{qq}=\sum_{k}\sigma _{k}^{q}\sigma
_{k}^{q},\;g^{qp}=\sum_{k}\sigma _{k}^{q}\sigma
_{k}^{p},\;g^{pp}=\sum_{k}\sigma _{k}^{p}\sigma _{k}^{p}$. We note that the corresponding Fokker-Planck equation \cite{Gardiner_SH} is
\begin{equation*}
\frac{\partial \varrho}{\partial t} =- \frac{\partial }{\partial q} \left( v^{q}\varrho \right)
-\frac{\partial }{\partial p} \left(v^{p} \varrho \right) 
+\frac{1}{2}\frac{\partial ^{2}}{\partial q^{2}} \left( g^{qq} \varrho \right)
+\frac{\partial ^{2}}{\partial q\partial p} \left( g^{qp} \varrho \right) 
+\frac{1}{2} \frac{\partial ^{2}}{\partial p^{2} } \left( g^{pp} \varrho \right) .
\end{equation*}

We observe that the definition of canonical flows (\ref{eq:Canonical}) makes sense in the stochastic setting.
From the It\={o} calculus \cite{Oksendal}, we now obtain the infinitesimal
form
\begin{equation}
d\left\{ f_t,g_t\right\} =\left\{ df_t,g_t\right\} +\left\{ f_t,dg_t\right\} +\left\{
df_t,dg_t \right\} ,  \label{eq:can_Ito}
\end{equation}
from which we find that the canonical condition is  
\begin{eqnarray*}
\left( \mathscr{D}_{\mathcal{L}}(f,g)-\sum_{k}\left\{ \sigma
_{k}\left( f\right) ,\sigma _{k}\left( g\right) \right\} \right) dt
+\sum_{k}\mathscr{D}_{\sigma _{k}.\nabla }(f,g)dQ_{k}\left( t\right) =0.
\end{eqnarray*}
To ensure that the $dQ_k (t)$ coefficient vanishes, we need to have $\mathscr{D}_{\sigma _{k}.\nabla }\equiv 0$
for each $k$ requiring that $\sigma _{k}.\nabla \equiv \left\{ \cdot
,F_{k}\right\} $, for some functions $F_{k}$, and so
\begin{equation*}
\mathscr{D}_{\mathcal{L}}(f,g)=\sum_{k}\big\{
\{f,F_{k}\},\{g,F_{k}\}\big\} .
\end{equation*}
It is easy to see \cite{Gough99} that the generator must then have the form
\begin{equation}
\mathcal{L}=\left\{ \cdot ,H\right\} +\frac{1}{2}\sum_{k}\big\{ \{\cdot
,F_{k}\},F_{k}\}\big\} .  \label{eq:Hormander}
\end{equation}
We now have 
\begin{eqnarray*}
v^{q} =\mathcal{L}(q) &=& \frac{\partial H}{\partial p}+\frac{1}{2}\sum_{k} \{  \frac{\partial F_{k}}{\partial p},F_k  \} ,\\
v^{p} =\mathcal{L}(p)&=&-\frac{\partial H}{\partial q}-\frac{1}{2}\sum_{k}\{ \frac{\partial F_{k}}{\partial q} , F_k \},
\end{eqnarray*}
so that the vector field $v$ (It\={o} drift) for canonical fields is then dissipative with
\begin{equation*}
\nabla .v=-\sum_{k}\left( \frac{\partial ^{2}F_{k}}{\partial q^{2}}\frac{%
\partial ^{2}F_{k}}{\partial p^{2}}-\left( \frac{\partial ^{2}F_{k}}{%
\partial q\partial p}\right) ^{2}\right) .
\end{equation*}
We summarize as follows.

\bigskip

\textbf{Theorem 1} \textit{The stochastic process on phase space defined by (%
\ref{eq:SDE}) is canonical with respect to the standard Poisson Brackets if
and only if the} $\sigma _{k}$ \textit{are Hamiltonian vector fields with
Hamiltonian} $F_{k}$ \textit{and} $\mathcal{L}$ \textit{takes the form}
(\ref{eq:Hormander}).

\bigskip

\textbf{Example 1:} We may try a linear function of $q$ and $p$ for the
fluctuation functions $F_{k}:$%
\begin{equation*}
F_{k}\left( q,p\right) =\alpha _{k}p+\beta _{k}q
\end{equation*}
for constants $\alpha _{k}$ and $\beta _{k}$. Unfortunately, this leads to
no dissipation since the It\={o} drift is Hamiltonian: $v^{q}=\frac{\partial H}{%
\partial p},v^{p}=-\frac{\partial H}{\partial q}$. This model does nothing
more than add Wiener noise onto an already Hamiltonian model. The
fluctuations have no dissipation to balance here.

\bigskip

\textbf{Example 2:} If we wish to obtain nonzero dissipation in the It\={o} drift, then
we need the $F_{k}$ to be nonlinear. The damped harmonic oscillator velocity
field  $v_{\mathrm{DHO}}$ (\ref{eq:DHO}) can be obtained from the single Wiener noise model with
the choices
\begin{eqnarray*}
H =\frac{p^{2}}{2m}+\frac{1}{2}m\omega ^{2}q^{2}+\frac{1}{2}\gamma qp, \quad
F =\sqrt{\gamma }\left( \frac{p^{2}}{2z }+\frac{1}{2}z q^{2}\right) .\, 
\end{eqnarray*}

\section{Symplectic Noise Models}

We start from the observation that in the
quantum case the noises come in canonically conjugate pairs $(Q_k(t),P_k(t)
) $, taken as dimensionless. We similarly postulate pairs of classical Wiener processes satisfying canonical Poisson Bracket relations
of the form
\begin{eqnarray}
\{ Q_j (t) ,P_k (s) \} = \frac{1}{ \mathsf{s} } \delta_{jk} \, \mathrm{min} (t,s) ,
\end{eqnarray}
where $\mathsf{s}>0$ has units of action, and then apply the same dynamical principles as before.

This is achieved by introducing the following definition of the \textit{full Poisson Brackets} for a classical system
with classical noise.

\bigskip

\textbf{Definition:} Let $(Q_k(t),P_k(t) )$, $k=1,\cdots,n$, be a family of $2n$ independent Wiener processes, and on the space of functions of the system variables $q,p$ \textit{and} the coordinate processes $(Q_k(t),P_k(t) )$ define the Poisson Brackets
\begin{equation}
\left\{ F ,G \right\} \triangleq \frac{ \partial F}{\partial q} \frac{ \partial G}{\partial p}-
 \frac{ \partial G}{\partial q}\frac{ \partial F}{\partial p}+
\frac{1}{ \mathsf{s} } \sum_k \int dt \left( \frac{ \delta F}{\delta Q_k (t) } \frac{ \delta G}{\delta P_k (t) }
- \frac{ \delta G}{\delta Q_k (t) } \frac{ \delta F}{\delta P_k (t) }  \right),
\label{eq:PB_full}
\end{equation}
for every pair of functionals $F,G$, with $\frac{ \delta F}{\delta Q_k (t) }$ denoting functional derivative, etc. 

\bigskip

A more concrete construction is given by taking $\mathcal{Q} (\cdot ), \mathcal{P} (\cdot )$ to be a pair of Gaussian random fields with moment generating functions
\begin{equation*}
\mathbb{E} \left[ e^{i \mathcal{Q} (f) + i \mathcal{P} (g) } \right] =
\exp \left\{  -\frac{1}{2} \int | f(t) |^2dt -\frac{1}{2} \int | g(t) |^2dt \right\}
\end{equation*}
and define the noise Poisson Brackets as the antisymmetric bi-derivation with basic relations
\begin{equation*}
 \left\{ \mathcal{Q} (f) , \mathcal{Q} (g) \right\}=0=\left\{ \mathcal{P} (f) , \mathcal{P} (g) \right\},\quad
\left\{ \mathcal{Q} (f) , \mathcal{P} (g) \right\} = \frac{1}{\mathsf{s}} \int f(t) g(t) \, dt ,
\end{equation*}
as well as satisfying a Jacobi identity. We then set $Q(t) = \mathcal{Q} ( 1_{[0,t]})$ and $P(t) = \mathcal{P} ( 1_{[0,t]})$, where $1_{[0,t]}$ 
is the indicator function of interval $[0,t]$.
Let $(X_t)$ and $(Y_t)$ be stochastic processes adapted to the filtration generated by the Weiner processes $Q,P$, then we have the infinitesimal relations
\begin{eqnarray*}
\{ X_t \, dQ (t) , Y_t \} = \{ X_t  , Y_t \}\, dQ (t), \: \{ X_t \, dP (t) , Y_t \} = \{ X_t  , Y_t \}\, dP (t), 
\end{eqnarray*}
and 
\begin{eqnarray*}
\{ X_t  dQ (t) , Y_t  dP(t) \} = \{ X_t  , Y_t \}\, dQ(t) dP(t) +X_tY_t \{ dQ(t) , dP(t) \} \equiv \frac{1}{\mathsf{s}} X_t Y_t \, dt .
\end{eqnarray*}

\subsection{Canonical Evolutions Under Symplectic Noise}

For a function $f$ of $q$ and $p$ we shall write $f_{t}$ for $f(q_{t},p_{t})$.
Let us consider a classical stochastic process of the form
\begin{eqnarray}
df_{t}=(\mathcal{L}f)_{t}\,dt+(\sigma .\nabla f)_{t}\,dQ(t) +(\varsigma .\nabla
f)_{t}\,dP (t)
\label{eq:SDE_can}
\end{eqnarray}
and we now require the stochastic flow to be canonical with respect to the
full Poisson Brackets ({\ref{eq:PB_full}). (For simplicity we consider a
single pair of canonically conjugate Wiener processes.) }

The equation (\ref{eq:can_Ito}) is required to hold with the full Poisson
Brackets ({\ref{eq:PB_full}) now understood. This time we have
\begin{eqnarray*}
\{df,dg\} &=&\{(\sigma .\nabla f)_{t}\,dQ+(\varsigma .\nabla f)_{t}\,dP, \\
&&(\sigma .\nabla g)_{t}\,dQ+(\varsigma .\nabla g)_{t}\,dP\} =\{\sigma .\nabla f,\sigma .\nabla g\}_{t}\,dt+\{\varsigma .\nabla
f,\varsigma .\nabla g\}_{t}\,dt \\
&&+ \mathsf{s}^{-1 } \left( \sigma .\nabla f)(\varsigma .\nabla g)_{t}dt-
 (\varsigma .\nabla
f)(\sigma .\nabla g)\right) _{t}\,dt ,
\end{eqnarray*}
and the equation (\ref{eq:can_Ito}) implies the following identities
\begin{eqnarray*}
\mathscr{D}_{\mathcal{L}}(f,g) &=&\left\{ (\sigma .\nabla f),(\sigma .\nabla
g)\right\} +\left\{ (\varsigma .\nabla f),(\varsigma .\nabla g)\right\}  \\
&&+ \mathsf{s}^{-1 } \left( \sigma .\nabla f)(\varsigma .\nabla g)- \mathsf{s}^{-1 }(\varsigma .\nabla
f)(\sigma .\nabla g\right)  ,\\
\mathscr{D}_{\sigma .\nabla }(f,g) &=&0,\\
\mathscr{D}_{\varsigma .\nabla
}(f,g) &=&0.
\end{eqnarray*}
Evidently, this requires that $\sigma .\nabla \equiv \{\cdot ,F\}$ and $%
\varsigma .\nabla \equiv \{\cdot ,G\}$, for some functions $F,G$ of $(q,p)$.
From arguments similar to before, we see that the generator takes the form
$\mathcal{L}=\{\cdot ,H\}+\frac{1}{2}\big\{ \{\cdot ,F\},F\big\} +\frac{1}{%
2}\big\{ \{\cdot ,G\},G\big\} +\mathcal{K}$,
where} $\mathscr{D}_{\mathcal{K}}(f,g)= \mathsf{s}^{-1 }\left( \sigma .\nabla f)(\varsigma
.\nabla g)- \mathsf{s}^{-1 } (\varsigma .\nabla f)(\sigma .\nabla g\right) $, and so the dissipation 
takes in the remarkable form
\begin{eqnarray*}
\mathscr{D}_{\mathcal{K}}(f,g)
&=&\frac{1}{\mathsf{s}} \{f,F\}\{g,G\}  -\{f,G\}\{g,F\} \\
&=& \frac{1}{\mathsf{s}} \left( \frac{\partial f}{\partial q}\frac{\partial F}{\partial p} -\frac{\partial F}{\partial q}\frac{\partial f}{\partial p}\right) \left( \frac{\partial g}{\partial q}\frac{\partial G}{\partial p}-\frac{\partial G}{\partial q}\frac{\partial g}{\partial p}\right) -\left( F\leftrightarrow
G\right)  \\
&\equiv& \frac{1}{\mathsf{s}} \{f,g\}\{F,G\}.
\end{eqnarray*}
Comparison with Proposition 1 shows that it is enough to take $\mathcal{K}$
to be a first order differential operator $u.\nabla$ with divergence $-\mathsf{s}^{-1 }\{F,G\}$. This
determines $u$ uniquely up to a Hamiltonian vector field which can always be
absorbed. We now synthesize the main result.

\bigskip

\textbf{Theorem 2} \textit{A diffusion on phase space driven by canonically
conjugate pairs of Wiener processes will be canonical for the full Poisson
Brackets } ({\ref{eq:PB_full}) \textit{if it takes the form}
\begin{equation*}
df_t = (\mathcal{L} f)_t \, dt + \sum_k \{ f, F_k \}_t \, dQ_k (t) + \{ f,
G_k \}_t \, dP_k (t)
\end{equation*}
\textit{with generator}
\begin{equation*}
\mathcal{L}= \{ \cdot ,H \} +\frac{1}{2} \sum_k \left( \big\{ \{ \cdot ,F_k \} ,F_k
\big\} + \big\{ \{ \cdot ,G_k \},G_k \big\} \right)+u.\nabla,
\end{equation*}
\textit{where $u$ is a vector field with $\nabla .u = -\mathsf{s}^{-1 } \sum_k\{ F_k
, G_k \}$, for functions $H,F_k,G_k$ on phase space.} }

Note that the dissipation $\mathscr{D}_{\mathcal{L}}(f,g)$ takes the general form
\begin{eqnarray}
 \sum_k \big\{ \{f,F_k\} ,\{ g ,F_k \} \big\} &&+ \sum_k \big\{ \{f,G_k\} ,\{ g ,G_k \} \big\} \nonumber \\
&& + \mathsf{s}^{-1} \sum_k\{F_k , G_k\} \,\{f,g\} .
\end{eqnarray}

The Fokker-Planck equation for the diffusion may be written in the form
\begin{equation*}
\frac{\partial \varrho }{\partial t}=\{ H, \varrho \} 
 +\frac{1}{2} \sum_k  \big\{ \{  \varrho , F_k \} , F_k \big\}
 +\frac{1}{2} \sum_k  \big\{ \{  \varrho , G_k \} , G_k \big\}
- \nabla . ( \varrho \, u ) .
\end{equation*}

\subsection{Linear Symplectic Stochastic Models}
We now shall describe general examples leading to a It\={o} drift velocity $v$ linear in the canonical coordinates, with bounded below Hamiltonian, and
constant diffusion matrix. For simplicity we restrict to a single canonical pair, and set 
\begin{eqnarray*}
H &=& H_0 +zqp, \\
H_0 &=&\frac{p^{2}}{2m}+\frac{1}{2}m\omega ^{2}q^{2}, \\
F &=&aq+bp, \\
G &=&cq+dp,
\end{eqnarray*}
and we set $\gamma =\mathsf{s}^{-1} \left\{ F,G\right\} =\left(
ad-bc\right) /\mathsf{s}$, which we assume to be strictly positive.

A particular solution is any vector field $u$ satisfying $\nabla .u=-\gamma $, and for definiteness we fix 
\begin{equation*}
u. \nabla =  -  \gamma  p \frac{\partial }{\partial p} .
\end{equation*}
The general solution is the particular solution plus some Hamiltonian vector field which we may absorb into $H$, and so 
we ignore in the present context.

This leads to the phase space velocity field

\begin{equation}
v^{q}=zq+\frac{1}{m}p, \quad  v^{p}=-m\omega
^{2}q-zp-\gamma p. 
\label{eq:v_z}
\end{equation}
The second-order part of the generator has the diffusion
tensor
\begin{equation*}
g =\left[ 
\begin{array}{cc}
b^{2}+d^{2} & ab+cd \\ 
ab+cd & a^{2}+c^{2}
\end{array}
\right] 
\end{equation*}
which is easily seen to be positive definite. Without loss of generality, let us fix on the choice
\begin{equation*}
F=-\sqrt{\gamma \mathsf{s}\epsilon }\,p,\quad G=\sqrt{\gamma \mathsf{s}/\epsilon }%
\,q,
\end{equation*}
where $\epsilon >0$ has units $\mathtt{m}\mathtt{N}^{-1}\mathtt{s}^{-1}$ 
(the general case follows from linear canonical transformations of the conjugate Wiener noises).

If we set $z=0$ then the system becomes invariant under position translation, and stochastic differential equations for the coordinate processes become
\begin{eqnarray}
dq_{t} &=&\dfrac{p_{t}}{m}\,dt-\sqrt{\gamma \mathsf{s}\epsilon }\,dQ(t), \nonumber  \\
dp_{t} &=&-(m\omega ^{2}q_{t}+\gamma p_{t})\,dt-\sqrt{\gamma \mathsf{s} /\epsilon }\,dP(t),
\label{eq:linearSDEs}
\end{eqnarray}
which are canonical with respect to the full Poisson Brackets (Theorem 2),
though not if we considered the system Poisson Brackets only (Theorem 1).  
The Fokker-Planck equation for the diffusion is
\begin{equation*}
\frac{\partial \varrho }{\partial t}=-\frac{\partial }{\partial q}\left( v_{%
\text{DHO}}^{q}\,\varrho \right) -\frac{\partial }{\partial p}\left( v_{%
\text{DHO}}^{p}\,\varrho \right) +\frac{1}{2}\gamma \mathsf{s}\epsilon \frac{%
\partial ^{2}\varrho }{\partial q^{2}}+\frac{1}{2}\frac{\gamma \mathsf{s}}{%
\epsilon }\frac{\partial ^{2}\varrho }{\partial p^{2}}
\end{equation*}
and  a steady state solution is given by a Gaussian
density, see \cite{Gardiner_SH} chapter 5, section 3,
\begin{equation*}
\varrho =Z^{-1}\exp \left\{ -\frac{1}{2}\left[ q,p\right] \sigma^{-1} \left[ 
\begin{array}{c}
q \\ 
p
\end{array}
\right] \right\} 
\end{equation*}
with covariance matrix $\sigma$ satisfying $A\sigma +\sigma A^\top =-g$ where $A=\left[ 
\begin{array}{cc}
0 & 1/m \\ 
-m\omega  & -\gamma 
\end{array}
\right] $. One finds that the Gaussian steady state is mean zero and
(positive definite!) covariance matrix $\sigma $ given by
\begin{equation*}
\left[ 
\begin{array}{cc}
\left\langle q^{2}\right\rangle_{\mathrm{ss}}  & \left\langle qp\right\rangle_{\mathrm{ss}}   \\ 
\left\langle pq\right\rangle_{\mathrm{ss}}   & \left\langle p^{2}\right\rangle_{\mathrm{ss}} 
\end{array}
\right] =\frac{1}{2}\mathsf{s}\left[ 
\begin{array}{cc}
\frac{1+\epsilon ^{2}m^{2}(\omega ^{2}+\gamma ^{2})}{\epsilon
^{2}m^{2}\omega ^{2}} & -\epsilon m\gamma  \\ 
-\epsilon m\gamma  & \frac{1+\epsilon ^{2}m^{2}\omega ^{2}}{\epsilon }
\end{array}
\right] .
\end{equation*}

It is customary to identify the quadratic function $\frac{1}{2}\left[ q,p \right] \sigma^{-1} \left[ 
\begin{array}{c}
q \\ 
p
\end{array}
\right] $ with $\beta H_{\text{eff}}\left( q,p\right) $ where $\beta $ is
the inverse temperature and $H_{\text{eff}}$ is the effective Hamiltonian.
However, it is clear that $ H_{\text{eff}}$ is not the oscillator Hamiltonian $H_0$. Indeed, we have cross terms in $H_{\text{eff}}$ due to the nonzero
covariances $\left\langle qp\right\rangle_{\mathrm{ss}}  =-\varepsilon m\gamma $, and so $%
H_{\text{eff}}\left( q,-p\right) \neq H_{\text{eff}}\left( q,p\right) $.

One may try to fix this by changing the Hamiltonian of the stochastic model
to $H=\frac{p^{2}}{2m}+\frac{1}{2}m\omega ^{2}q^{2}+zqp$ for some real
scalar $z$, in which case one computes that
\begin{eqnarray*}
\left\langle q^{2}\right\rangle_{\mathrm{ss}}  &=& \frac{1}{Y(z)} \left( 2\varepsilon ^{2}m^{2}\omega ^{2}+2\varepsilon ^{2}m^{2}z^{2}+4\varepsilon
^{2}m^{2}\gamma z+2\varepsilon ^{2}m^{2}\gamma ^{2}+2 \right) \nonumber \\
\left\langle p^{2}\right\rangle_{\mathrm{ss}}  &=& \frac{1}{Y(z)} \left(
2m^{4}\omega ^{4}\varepsilon ^{2}-m^{2}z^{2}-m^{2}\gamma
z+2m^{2}\omega ^{2}
\right) \nonumber \\
 \left\langle qp\right\rangle_{\mathrm{ss}} &=& - \frac{1}{Y(z)} \left( 2m^{3}\varepsilon
^{2}\omega ^{2}z + 2m^{3}\varepsilon ^{2}\omega ^{2}\gamma + mz\right)
\end{eqnarray*}
where $Y(z) = \frac{2 m^2}{\gamma \mathsf{s}}  \left( z\omega
^{2}-z^{3}-2z^{2}\gamma -z\gamma ^{2}+2\omega ^{2}\gamma \right)$.
The specific choice $z=-2\varepsilon ^{2}m^{2}\omega ^{2}\frac{\gamma }{%
2\varepsilon ^{2}m^{2}\omega ^{2}+1}$ leads to steady state $\left\langle
qp\right\rangle_{\mathrm{ss}} =0$, that is

\begin{equation*}
\sigma =\frac{1}{2} \mathsf{s}  \frac{2\varepsilon ^{2}m^{2}\omega ^{2}+1}{\varepsilon
m^{2}\omega ^{2}}\left[ 
\begin{array}{cc}
1 & 0 \\ 
0 & m^{2}\omega ^{2}
\end{array}
\right] .
\end{equation*}
This is the Gibbs state giving the canonical ensemble for the energy $H_0$ with inverse temperature $\beta =1/k_B T$ given by
\begin{equation*}
k_B T = \frac{1}{2} \mathsf{s} \frac{2 \epsilon^2 m^2 \omega^2 +1}{ \epsilon m},
\end{equation*}
and we have a proper equipartition of energy.
However, the underlying Hamiltonian is now have $H\left( q,p\right) = H_0 \left( q,p\right) +zqp$ and the
stochastic differential equations no longer equals (\ref{eq:linearSDEs}) since the drift velocity is now
given by (\ref{eq:v_z}) but with the non-zero value of $z$ quoted above.

\section{Comparison With Quantum Models}

It is instructive to consider the quantum Markov of a linearly damped harmonic oscillator. 
For linear damped motion for an open system with canonical
observables operators $q,p$ satisfying the Heisenberg commutation relations $[q,p]=i \hbar$, the most general form can be obtained by taking $H= H_0 +
\frac{\mu}{2} ( qp+pq), \; L_k = \alpha_k \, p + \beta_k \, q $, where $H_0
= \frac{p^2}{2m} + \frac{1}{2} m \omega^2 q^2$ is a harmonic oscillator
Hamiltonian, $\mu$ is a real constant, the $\alpha_k, \beta_k$ are complex
constants, and it suffices to take just two coupling terms $k=1,2$ \cite
{Lind,Barchielli,ISS}. We follow the presentation given in \cite{Vacchini}. 

For instance, let us, introduce the annihilator $a= \sqrt{ m\omega / 2\hbar }(q+ i p/ m\omega )$ so that $H= \hbar \omega (a^\dag a + \frac{1}{2}
)$, and take a pair of independent quantum Wiener processes $B_k (t)$  ($k=1,2$) with coupling operators $L_1= \sqrt{\gamma (n+1)} a$
and $L_2= -\sqrt{\gamma  n} a^\dag$, leads to
\begin{eqnarray*}
dq_t &=&  \left( \frac{p_t}{m} + ( \frac{1}{2} \gamma -\mu )q_t \right) \, dt 
- \sqrt{\frac{\gamma (n+1) \hbar }{2 m \omega}} \, dQ_1 (t) 
- \sqrt{\frac{\gamma n \hbar }{2 m \omega}} \, dQ_2 (t )
,\\ 
dp_t &=&  \left( -  m\omega ^{2}q_t-( \frac{1}{2} \gamma + \mu ) p_t \right) \, dt \\
&& -
\sqrt{\frac{\gamma (n+1) \hbar m \omega}{2}} \, dP_1 (t)
+
\sqrt{\frac{\gamma n \hbar m \omega}{2}} \, dP_2 (t),
\end{eqnarray*}
where $q_t =j_t (q)$, $p_t =j_t (p)$ and the $Q_k , P_k$ are the quadrature processes.

If we take $\mu=\gamma$ and set $n=0$ (vacuum noise) then the Langevin equations are identical in form to
the classical symplectic SDEs (\ref{eq:linearSDEs}) above under the identification $\mathsf{s}=\hbar /2$ and $\varepsilon =1/m\omega $.
However, once again we do not have convergence a Gibbs state of the form $e^{-\beta H_0}/Z$, with $H_0$ the oscillator Hamiltonian.

If, however, we set $\mu=0$ and $n>0$ then we find convergence to the Gibbs state $e^{-\beta H_0}/Z$ with temperature determined by
$n= 1/( e^{\beta \hbar \omega } -1)$. (The same actually applies to the vacuum case $n=0$ where the oscillator decays to its ground state as equilibrium state.) 
However, the drift velocity in phase space is now $v^q = \frac{p}{m} +  \frac{1}{2} \gamma q$,
$v^p = -m \omega^2 q -  \frac{1}{2} \gamma p$, and this is derived from the Hamiltonian $H= H_0 +
\frac{\gamma}{2} ( qp+pq)$, as opposed to the vector field $v_{\mathrm{DHO}}$ of linearly damped harmonic oscillator.

\section{Conclusion}
The Poisson Brackets are the algebraic ghosts of quantum mechanics in the classical world. The Hamiltonian and Poisson bracket formulation of classical mechanics 
dates from the 1830's, and it was Dirac who realised that the commutator $\frac{1}{i\hbar} [ \cdot , \cdot ]$ is the analog, in spirit, of the Poisson Brackets.
It has become fashionable to view quantum probability as the natural extension of classical theory that formulates the classical theory axiomatically, and then 
drops the assumption of commutativity of random variables. While this has indeed been productive, it ignores the fact that the specific form of the non-commutativity - the Heisenberg commutation relations - have a physical importance that transcends the abstract formalism. 

The new ingredient introduced in this paper is the idea of canonically conjugate noises, and more generally of a Poisson Brackets for the noise in stochastic classical mechanical models. The idea is not unreasonable as it proposes that a classical environment is ultimately a mechanical system, and that an idealized model of the noise should somehow retain the symplectic structure. There is a central program in Mathematical and Statistical Physics to realise quantum models describing dissipation and relaxation to thermal equilibrium. Here there is two-way transfer from quantising classical stochastic models and dequantising quantum open systems back to the classical world. We in effect provide a missing link - a classical noise which its own inherent symplectic structure where the physical noises do not Poisson-commute!

If we take the noise to have no symplectic nature, but require the diffusion on phase space to be canonical for the system Poisson Bracket, then we arrive at the 
general situation described in Theorem 1, which was previously derived in \cite{Gough99}, see also \cite{Sinha}. Here we have the surprising fact that to dilate the equations of motion 
of the linearly damped Harmonic oscillator (\ref{eq:DHO}) that we take a quadratic Hamiltonian $H$ but must have a coupling $F$ to the noise that is \textit{quadratic}.
In the quantum case it suffices to take the coupling operators $L_k$ to be \textit{linear} in the canonical observables. Indeed, we see in Example 1 that linear $F_k$ lead to zero dissipation.

If we now allow the noise to have a symplectic structure, and ask for the overall system plus noise Poisson brackets to be respected, then the general situation is as given in Theorem 
2. In the quantum model, the system interacts with the noise, so that in the Heisenberg picture observables of the system evolve as stochastic process on the tensor product of the system and noise Hilbert spaces. When we use the commutator in the quantum stochastic setting we really do have the commutator of operators on the Hilbert space of the system tensored with the Fock space of the noise. In that sense, our use of the full Poisson Brackets for system plus noise capture the spirit of what is going on in the quantum models.

The $q$ and $p$ play the roles of the $\alpha$ and $\beta$ variables in the Onsager-Casimir-Machlup theory. 
A standing assumption of Onsager and Machlup \cite{Onsager_Machlup}, in the case of Gaussian fluctuations for the $\alpha = q$ and $\beta = p$ variables,
is that there are no covariances between the $q$'s and $p$'s, consistent with the intuition that the entropy cannot change under time reversal. Interpreting the state as a Gibbs state
means that we have a quadratic Hamiltonian satisfying $H(q,-p)=H(q,p)$, so there can be no $q$ and $p$ cross terms.  
We propose that the stochastic flows of the form described in Theorem 2 give the appropriate noise model on which to base the theory of dissipation and 
spontaneous fluctuations about equilibrium.

Inevitably, whenever we consider Wiener noise we are making an approximation to the physical spectrum of frequencies by the flat spectrum of white noise. However, we should note that the approximation is an excellent and widely accepted one in quantum optics, and what we gain is 
computationally tractable model that allows us to use the It\={o} calculus. Noting that the model noise is just a mathematical proxy for the physical environment, what we have 
tried to do is to retain the canonical nature of the environment in the noise model \cite{ALV,Gough_limit}. This is automatic in the quantum theory, but classically we have to define a Poisson structure for the noise, along with the notion of canonically conjugate Wiener processes.

Within the framework of Theorem 2, we now see that it is possible to have a linear noise coupling leading to a stochastic diffusion on phase space with the linearly damped Harmonic oscillator (\ref{eq:DHO}) as drift. This much agrees with the quantum situation. Moreover, if we wish to have a linear model with constant damping, then we cannot have both the drift given by (\ref{eq:DHO}) and a steady state that is not a thermal equilibrium state for the oscillator. This is well-known in the quantum setting, where it has been much touted as the failure of the quantum regression theorem. This is frequently blamed on the structure of GKS-Lindblad generator, which is in turn attributed to either the quantum Markov assumption or complete positivity, however, we see here that virtually the same situation occurs classically and that the real transgressor is the requirement of a symplectic model for both system and noise.

\begin{acknowledgements}
The author has the pleasant duty to thank Dr Guofeng Zhang for the kind hospitality
shown during his visit to the Hong Kong Polytechnic University in Winter 2014 during
which time this paper was written, and to the Royal Academy of Engineering for supporting this
this through their UK-China exchange scheme.
\end{acknowledgements}

\begin{thebibliography}{}
%
%

\bibitem{Onsager}
Onsager L.: Reciprocal Relations in Irreversible Processes, 
Phys. Rev. \textbf{37}, 405-426 (1931), and \textbf{38}, 2265-2279 (1931)

\bibitem{Casimir}
Casimir H.B.G.: On Onsager's Principle of Microscopic Reversibility, Revs. Modern Physics, \textbf{17}, 343, (1945)


\bibitem{Onsager_Machlup}
Onsager, L., Machlup, S.: Fluctuations and Irreversible Processes
Phys. Rev. \textbf{91}, 1505-1512 (1953);
and, Fluctuations and Irreversible Processes II, Phys. Rev. \textbf{91}, 1512-1515 (1953)

\bibitem{Nyquist}  Nyquist, H.: Thermal Agitation of Electric Charge in Conductors, Phys. Rev. \textbf{32}, 110 (1928);  Callen, H.B., Welton, T.A.:
Irreversibility and Generalized Noise, Phys. Rev. \textbf{83}, 34 (1951); Kubo, R.: Statistical-mechanical theory of irreversible processes. 1. general theory 
and simple applications to magnetic and conduction problems, J. Phys. Soc. Japan \textbf{12}, 570 (1954).

\bibitem{Talkner} P. Talkner, On The Failure of the Quatum Regression Hypothesis,  Ann. Phys., \textbf{167}, 390 (1986)

\bibitem{GKS} V. Gorini, A. Kossakowski, E.C.G. Sudarshan, Completely positive semigroups of N-level systems.
J. Math. Phys. \textbf{17} (5): 821-825 (1976)

\bibitem{Lindblad} Lindblad, G.: On the generators of quantum dynamical semigroups, Commun. Math. Phys. \textbf{48}, 119 (1976)

\bibitem{HP}  Hudson, R., and Parthasarathy, K.R.:
Quantum Ito's formula and stochastic evolution,
Commun. Math. Phys. \textbf{93}, 301-232 (1984)


\bibitem{Partha} Parthasarathy, K.R.:  \textit{An introduction to quantum
stochastic calculus}, (Birkhauser, Berlin, 1992)

\bibitem{ALV} Accardi, L., Lu, Y.G., Volovich,I.:  \textit{Quantum theory and its stochastic limit}, Springer, Berlin Heidelberg (2002)

\bibitem{GZ}  Gardiner, C.W. and  Zoller, P.: \textit{Quantum Noise},
Springer-Verlag, Berlin and New York, 3rd edition, (2004)

\bibitem{KM} K\"{u}mmerer, B., Maassen, J.: The essentially commutative dilations of dynamical semigroups on $ M_n$.
Comm. Math. Phys. \textbf{109}, no. 1, 1-22 (1987)

\bibitem{BGL} Bakry, D., Gentil, I. and Ledoux, M.: \emph{Analysis and Geometry of Markov Diffusion Operators}, Grundlehren der mathematischen Wissenschaften Vol. 348,
Springer (2014)

\bibitem{Bloch} Bloch, A.: \textit{Nonholonomic Mechanics and Control: With the Collaboration of J.Baillieul, P.Crouch and J.Marsden} (Interdisciplinary Applied Mathematics),
Springer, New York (2003)

\bibitem{Gough99}  Gough, J.E.: Dissipative canonical flows in classical and quantum mechanics, 
Journ. of Math. Physics, \textbf{40}, Issue 6,
pp. 2805-2815 (1999)

\bibitem{Sinha} Sinha, K.B.: Quantum mechanics of dissipative systems, J. Ind. Inst. Sci. 77, 275–279 (1997)

\bibitem{Oksendal} Oksendal, B.: \textit{Stochastic Differential Equations:
An Introduction with Applications}, Universitext, Springer; 6th edition
(2010)

\bibitem{Vacchini} Vacchini, B.: Quantum optical versus quantum Brownian motion master equation in terms of covariance and equilibrium properties, 
Journ. Math. Physics \textbf{43}, 5446-5458 (2002)

\bibitem{Lind}  Lindblad, G.: Brownian motion of a quantum harmonic oscillator, Rep. Math. Phys. \textbf{10}, 393 (1976)

\bibitem{Barchielli}  Barchielli, A.: Continual measurements for quantum open systems, Nuovo Cimento Soc. Ital. Fis., B 74B,
113 (1983)

\bibitem{ISS}  Isar, A., S\v{a}ndulescu, A., Scutaru, H., Stefanescu, E. and Scheid, W.: Open quantum Systems, Int. J. Mod. Phys. E \textbf{3}, 635 (1994)

\bibitem{Gardiner_SH} Gardiner, C.W.: \textit{Handbook of Stochastic Methods: For Physics, Chemistry and Natural Sciences} (Springer Series in Synergetics)
Springer, Berlin Heidelberg (1985)

\bibitem{Gough_limit} Gough, J.: Asymptotic stochastic transformations for nonlinear quantum dynamical systems
J Gough, Rep. on Math. Phys. \textbf{44} (3), 313-338	16	(1999); and, Noncommutative itô and stratonovich noise and stochastic evolutionsand stratonovich noise and stochastic evolutions
Theoret. and Math. Phys. \textbf{113} (2), 1431-1437	16	(1997); and,	Causal structure of quantum stochastic integrators, Theor. and Math. Phys. 111 (2), 563-575,(1997); and, Quantum Stratonovich calculus and the quantum Wong-Zakai theorem, Jour. Math. Phys. \textbf{47}, 11, 113509, (2005)

\end{thebibliography}


\end{document}